\documentclass{article}
\usepackage{spconf,amsmath,graphicx,color}
\usepackage{booktabs, multirow} 
\usepackage{array, makecell}
\usepackage{caption} 
\usepackage{soul}
\usepackage[table]{xcolor} 
\usepackage[nolist]{acronym}
\usepackage{changepage,threeparttable} 

\newacro{MAC}[MAC]{multiplication addition count}
\newacro{FLOPs}[FLOPs]{floating point operations}
\newacro{SED}[SED]{sound event detection}
\newacro{KD}[KD]{knowledge distillation}
\newacro{NAS}[NAS]{neural architecture search}
\newacro{SI-SDR}[SI-SDR]{scale-invariant signal to distortion ratio}
\newacro{MCUs}[MCUs]{micro-controller units}
\newacro{MIPS}[MIPS]{milion instructions per second}
\newacro{BLSTM}[BLSTM]{bi-directional long short term memory}

\title{Scaling strategies for on-device low-complexity \\ source separation with Conv-Tasnet}
\name{}

\twoauthors
  {Mohamed Nabih Ali}
	{University of Trento, Trento, Italy\\{\tt mohamed.nabih@unitn.it}}
  {Francesco Paissan, Daniele Falavigna, Alessio Brutti}
	{Fondazione Bruno Kessler, Trento, Italy\\{\tt (fpaissan,falavi,brutti)@fbk.eu}}
	
\begin{document}
\ninept

\maketitle

\begin{abstract}
Recently, several very effective neural approaches for single-channel speech separation have been presented in the literature. However, due to the size and complexity of these models, their use on low-resource devices, e.g. for hearing aids, and earphones, is still a challenge and established solutions are not available yet. Although approaches based on either pruning or compressing neural models have been proposed, the design of a model architecture suitable for a certain application domain often requires heuristic procedures 
not easily portable to different low-resource platforms.
Given the modular nature of the well-known Conv-Tasnet speech separation architecture, in this paper we consider three parameters that directly control the overall size of the model, namely: the number of residual blocks, the number of repetitions of the separation blocks and the number of channels in the depth-wise convolutions, and experimentally evaluate how they affect the  speech separation performance. In particular, experiments carried out on the Libri2Mix show that the number of dilated 1D-Conv blocks is the most critical parameter and that the usage of extra-dilation in the residual blocks allows reducing the performance drop.

\end{abstract}
\section{Introduction}
\label{sec:intro}
Today a variety of applications successfully employ speech technology features to improve the overall end-user satisfaction \cite{li2015robust,Jayashri_2016}. Although robust solutions and commercial products exist for many applications domains, several performance limitations  are still present, in particular when the environment and the noise conditions are less controlled. One open challenge is the capability to handle scenarios where overlapping speech occurs \cite{Lu_2021,cornell2022overlapped}. In particular thanks to the use of neural models, recently single-channel speech separation has steadily progressed, achieving extremely high separation performance. 
Despite their efficacy, these models generally require estimating a large number of parameters, thus demanding noticeable computational and memory 
resources that prevent their usage on devices with limited capabilities, e.g. hearing aids or \ac{MCUs}. Table~\ref{tab:sota} reports the characteristics of some of the currently most popular separation approaches in the literature in terms of number of parameters and \ac{FLOPs} during inference. Note that all models require MB of RAM to run, even assuming that quantization is applied, with Dual-Path-RNN particularly eager for what concerns computation. 
\begin{table}[!htp]\centering
\begin{small}
\caption{Recent neural architectures for source separation, alongside the overall number of parameters (in millions) and Giga-\ac{FLOPs} in inference on one second of input audio 
\cite{tzinis2020sudo}.}
\label{tab:sota}
\begin{tabular}{l|cc}\toprule
{Method} &  {\# Param.(M)} &{FLOPs (G)} \\
\hline
Two-Step TDCN \cite{tzinis2020two} &8.63 &7.09\\%
Conv-Tasnet \cite{luo2019conv} &5.10 &5.23\\%
Dual-Path-RNN \cite{luo2020dual} &2.63 &48.89\\%
SuDoRM-RF \cite{tzinis2020sudo} &2.66 &2.52\\
\toprule
\end{tabular}
\end{small}
\end{table}
To contextualize the numbers above, Table~\ref{tab:platforms} reports some examples of current processing platforms ranging from low-end \ac{MCUs} to single-board computers\footnote{An exact mapping between available \ac{MIPS} and \ac{FLOPs} is not possible since it depends on the firmware and the actual implementation of the model}. 
\begin{table}[th!]
\centering
 \caption{Examples of embedded platforms and their hardware capabilities in terms of RAM and \ac{MIPS}.}

\begin{tabular}{l|ll}
\toprule
Board Name             & RAM{[}KB{]}  & \ac{MIPS}\\ \hline
STM32L476RG            & 128         & 80       \\ 
TI MSP432P4111         & 256         & 58.56    \\ 
BeagleBone Black       & 524288      & 1607     \\
Raspberry Pi 3 B+      & 1048576      & 2800    \\
\hline
\end{tabular}
\label{tab:platforms}
\end{table}

The SOTA models reported in Table~\ref{tab:sota} are clearly not suitable to operate on MCUs like the STM32L476RG or TI MSP432P4111, which features few hundreds KB of RAM. Their deployment would be problematic also on more powerful devices like BeagleBone Black and Raspberry Pi. If the available memory and computational capacity would be sufficient to run some of the models reported in Table~\ref{tab:sota}, such as SuDoRM-RF and possibly Conv-Tas-Net, they would not allow any other service to be operational on the device.

This issue is common to several applications relying on neural models. As a consequence, a plethora of approaches have been published towards fitting large neural networks into resource-constrained devices: quantization~\cite{yang2019quantization}, pruning~\cite{valerio2020dynamic}, model compression via \ac{KD}~\cite{hinton2015distilling}, \ac{NAS}~\cite{Mo_2021}, and so forth. Although these methods have been successfully used in a variety of tasks and domains, one limitation is that they are not scalable and they still require a specific design and manual efforts to obtain a reduced model that fits the requirement of a given resource-constrained device or, more generally, the resource budget allocated to the task.at hand. Conversely, a scalable model that can be adjusted with few parameters is very appealing from an application point of view.

A widely used separation model in recent scientific literature is Conv-Tasnet \cite{luo2019conv}. One of its relevant features is its modular nature that allows  to easily scale down the model based on the available resources. Basically, the model's computational complexity is controlled by three parameters: the number of dilated residual blocks, the number of repetitions of the separation blocks, and the number of channels in the depth-wise convolutions. The main contribution of this work consists in measuring and analyzing how the three parameters mentioned above affect the overall separation performance, thus providing useful suggestions for future implementation of the neural model on smart devices. We have experimentally observed that the number of residual blocks is a crucial parameter, since it determines the maximum dilation factor and, as a consequence, the size of the receptive fields. We have also observed that increasing the dilation step noticeably limits the performance drop when reducing the number of blocks.

This paper is organized as follows: section~\ref{sec:related} summarizes some scientific papers  related to the work here described, section~\ref{sec:approach} describes the method proposed for scaling the complexity of the Conv-Tasnet model, section ~\ref{sec:experiments} reports and discuss the results and section~\ref{sec:page} concludes the paper.

\section{Related Work}
\label{sec:related}
Recently, single channel sound source separation has been  approached with deep learning  \cite{weninger_2014, wang_2015, erdogan_2015}. In particular, prior architectures based on encoder, decoder and separation modules have been effectively improved with the introduction of end-to-end neural models, such as the aforementioned Conv-Tasnet, where all the needed processing is performed in a single stage by optimizing a loss that minimizes the distortion between the model output and a clean target signal.
However, deploying these models on edge devices, e.g. in embedded applications or smart  mobile, portable devices, is a challenging task due to the computational complexity of current models, which leads to extensive memory and computation usage. Therefore, several studies have been focused on developing neural modules and architectures demanding small computational resources. 
One popular example are the depth-wise separable (DWS) convolutions~\cite{sifre2014rigid, chollet2017xception} that allows to implement convolution operations, i.e. the basic transformations employed in audio and video neural processing solution, with a reduced number of arithmetical operations. 

Regarding the speech domain, DWS were applied in monaural singing voice separation in \cite{pyykkonen2020depthwise}. 
%
Similar strategies are used in Mobile-Nets, a family of light-weight deep neural networks proposed in \cite{howard2017mobilenets} for edge-devices. However, the authors highlight that a large dilation factor in separable convolutions often introduces several artifacts. Recently, more  studies propose meta-learning algorithms to optimize architecture configurations for efficient computational resources with promising performance \cite{yu2019universally,cai2019once}.

Regarding the research specific for the speech domain, \cite{kalchbrenner2018efficient} describes "WaveRNN", a  model for addressing  speech synthesis tasks. The motivation behind its usage is to lower the time to generate synthesised samples with an acceptable  quality. This compact model allows to generate a 24 kHz 16-bit audio 4× faster than real time on a GPU. The authors propose also to use 
 a pruning technique that reduces significantly the number of the model weights, thus achieving excellent  performance in terms of \ac{FLOPs} and latency.  
 
 The authors of \cite{jeon2020lightweight} propose a light-weight model based on the U-Net \cite{stoller2018wave} architecture for effective speech separation on edge computing devices. To do this the inception-like multi-lane dimensionality reduction technique is utilized in each convolutional layer. 
 
 In group communication (GroupComm), proposed in \cite{luo2021ultra},
 an $N$-dimensional input feature vector is split into $K$ groups, each represented by a small feature vector of dimension $M$. A small separation module is then shared by all the groups. With this approach the model achieves performance similar to those of the original dual-path RNN model but with 
 35.6x fewer parameters, and with 2.3x fewer MAC operations.
 
 Two steps light-weight online speech separation was proposed in \cite{maldonado2020lightweight}. Firstly, a phase-based beamformer was used to estimate the source of interest. Then, a time-frequency binary mask estimator, based on \ac{BLSTM} cells, was used to calculate the binary mask for each one of the sources. 
 
Finally, in \cite{chen2018distilled}, the authors applied two steps model compression based on binarization and distillation approaches: the model is firstly binarized, and then knowledge distillation is used during the training stage to improve the separation performance.

Note that the approaches specifically designed for the source separation problem are limited in number. In addition, all of them are tailored to specific resources and are not easily adjustable to different devices. Instead, the analysis  described in the next section allows us to determine the optimal network architecture that, given the resources available on device, minimizes the performance drop. 


\section{Scaling Approach}
 \label{sec:approach}
\begin{figure}[htbp]
\centering
\includegraphics[width=8cm]{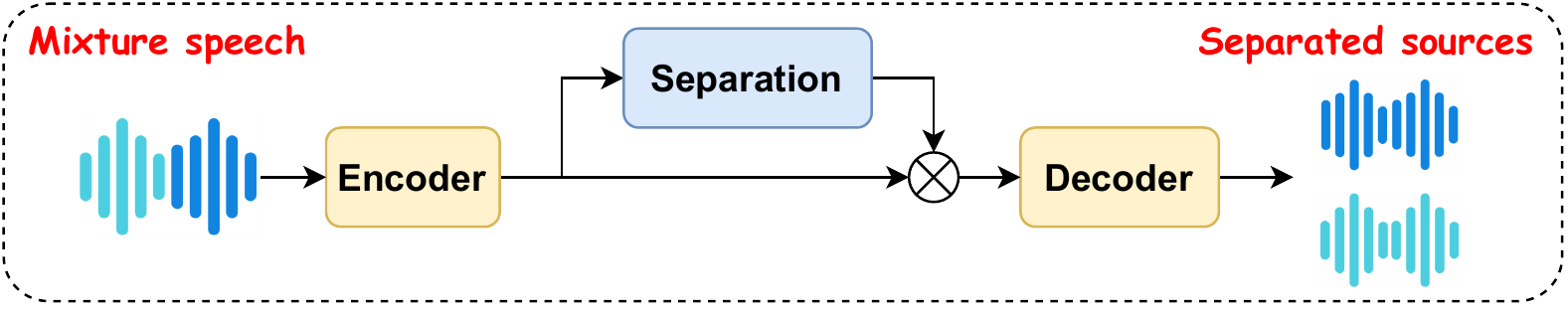}
\caption{Conceptual block diagram of the Conv-TAS-Net architecture}
\label{fig:convtasnet}
\end{figure}

\begin{figure*}[htbp]
    \centering
    \includegraphics[width=0.7\textwidth]{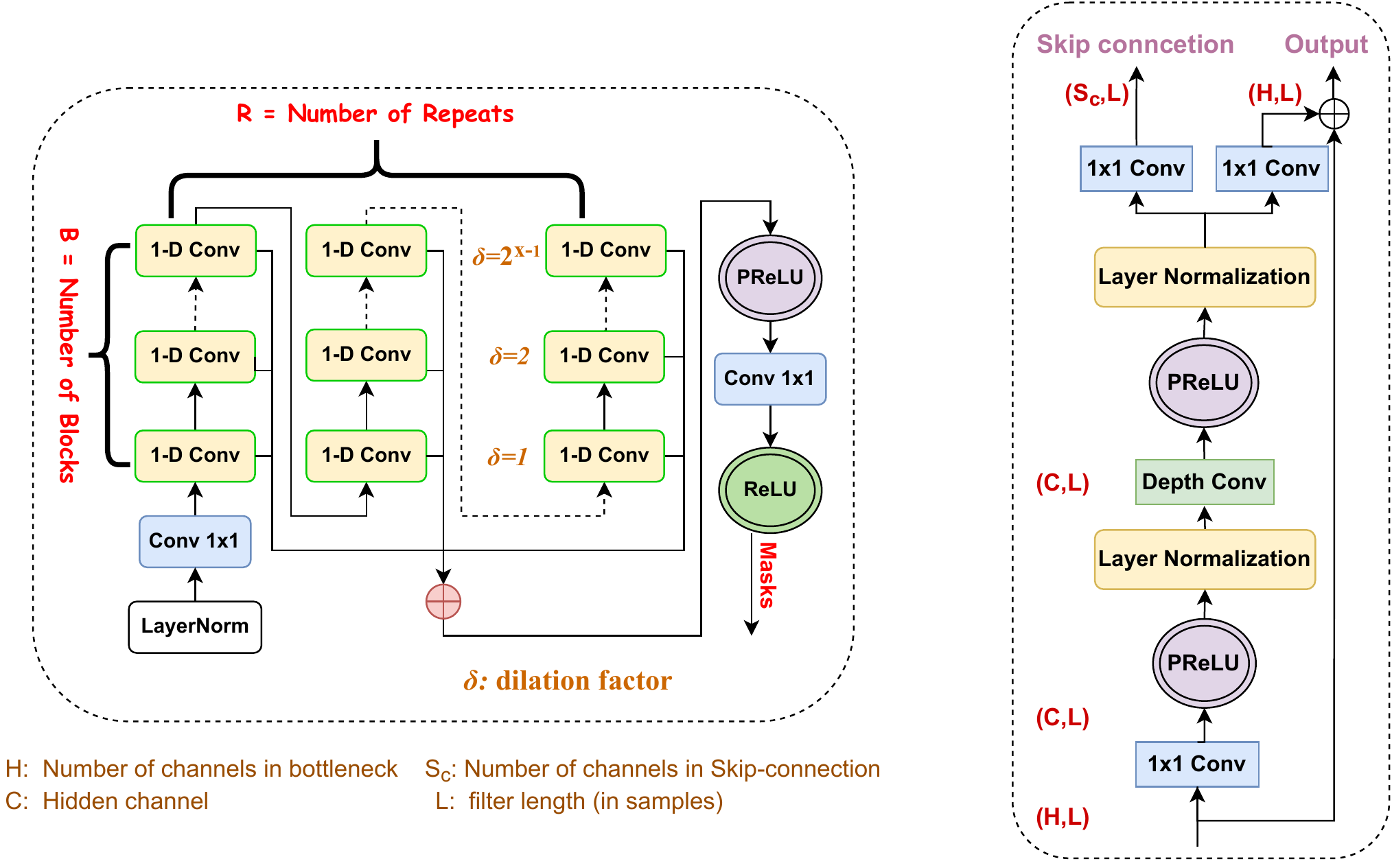}
    \caption{Block diagram of the Conv-Tasnet separation block (left) and of a single 1-D dilated convolutional block (right)}
    \label{fig:separation}
\end{figure*}

Figure~\ref{fig:convtasnet} shows the conceptual block diagram of the Conv-Tasnet architecture~\cite{luo2019conv}; it consists of 3 blocks: encoder, separation, and decoder. In this work, we focus on the separation block which is the most computationally eager part. We do not consider a reduction of the dimensions of the encoded features as it would considerably affect the quality of the signal reconstruction by the decoder even in presence of an accurate separation. The separator block is depicted in Figure~\ref{fig:separation}: it includes a sequence of 1-D dilated residual convolutional blocks followed by a mask generator. The modular nature of the separation block allows one to easily scale the model given a target memory occupancy and \ac{MAC} or \ac{FLOPs}. 

In particular, Figure~\ref{fig:separation} highlights the 3 scaling parameters taken into account in our analysis, that is: {\it i)} the number of dilated blocks $B$; {\it ii)} the number of repetitions of the dilated block sequences $R$ and {\it iii)} and the number of channels $C$ employed in the depth-wise convolution in each residual block (see the right part of Figure~\ref{fig:separation}). The reduction of the latter, in particular, leads to the use of inverted residual convolutional blocks as successfully applied in~\cite{howard2017mobilenets, Paissan_2022}. The values of these parameters in the baseline model are: $B=8$, $R=3$ and $C=512$. As shown in the figure, the dilation in the residual blocks is exponentially increased as $\delta=2^{\left(i-1\right)}$, where $i$ is the index of the residual block ($i=1,\dots,B$). Therefore, the value of $B$ is particularly crucial as it determines the maximum dilation applied in the model. In line with our previous work on PhiNets applied to \ac{SED} ~\cite{Brutti-22}, we observe that similar memory footprints and \ac{FLOPs} can be achieved with different configurations of the three scaling parameters, that result in largely different separation performance.

\section{Experimental Analysis and Results}
\label{sec:experiments}
Our experiments are based on models and recipes provided in the Asteroid toolkit~\cite{Pariente2020Asteroid} and have been carried out on the Libri2Mix dataset~\cite{cosentino2020librimix}, obtained with the aforementioned toolkit. We consider the "train-100" training set, the "sep-clean" task, i.e. the input signal is a mixture of two-speaker clean sources without noise, and mode "min", i.e. the mixture signal ends with the shortest source. We use the training parameters defined in the repository: learning rate $10^{-3}$, 200 epochs with early stopping, patience 30, and batches with 24 segments. As a common practice in the source separation community, performance is measured in terms of \ac{SI-SDR}~\cite{Vincent-2006}. 

As mentioned above, we performed experiments by varying: the number $B$ of residual blocks in the Conv-Tasnet model: $B=2,4,6,8$; the number $R$ of repeated blocks ($R=1,2,3$); the number of channels in the depth-wise 1D-Conv blocks ($C=512,128,64$). 

\begin{figure}
    \centering
    \includegraphics[width=9cm]{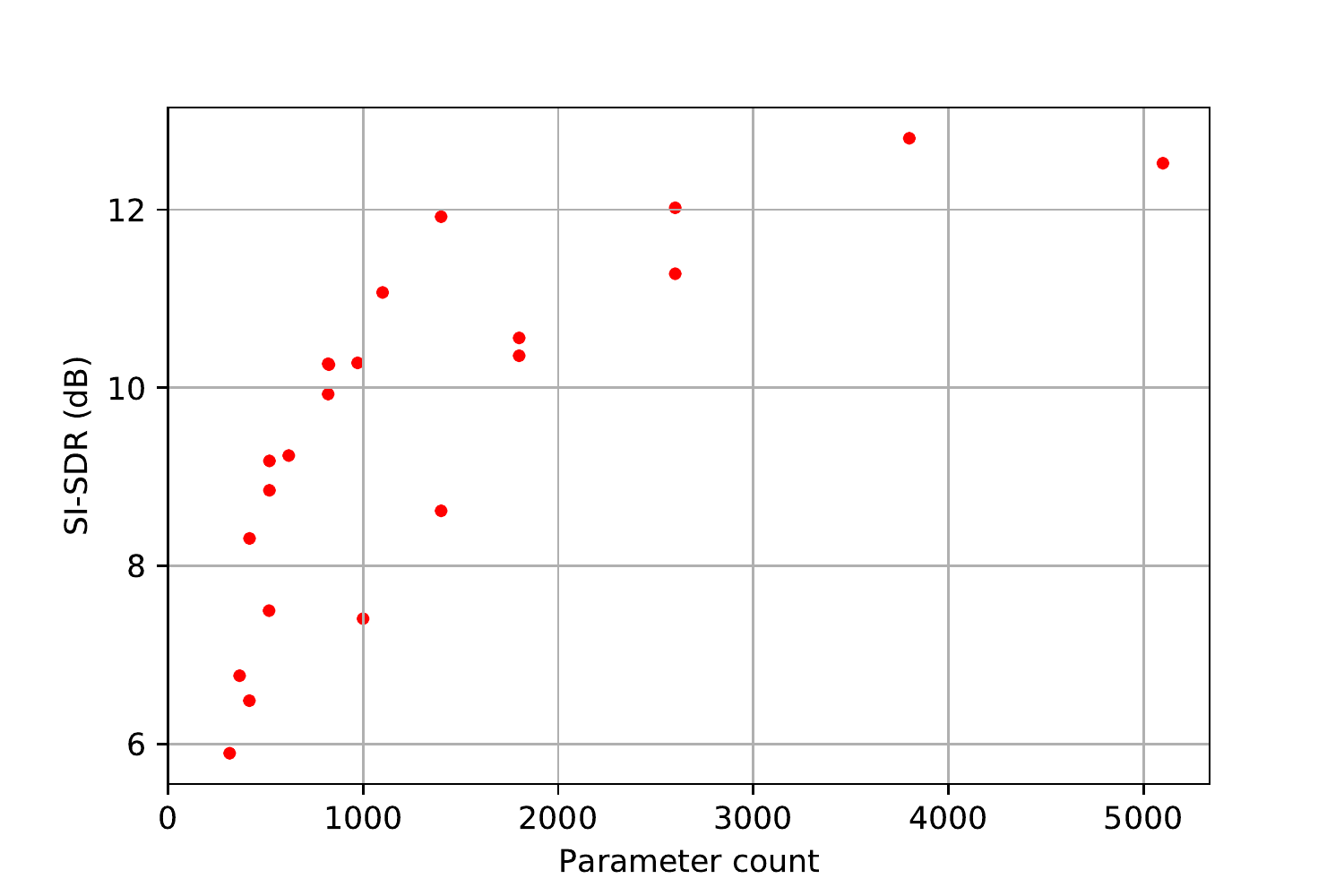}
    \caption{SI-SDR provided by separation models with varying number of parameters}
    \label{fig:sdr-parameters}
\vspace{-0.8cm}
\end{figure}

Figure~\ref{fig:sdr-parameters} reports the performance of each configuration as a function of the number of parameters (parameter counts) in each corresponding model. We notice, as expected, that a reduction in the number of model parameters determines a decrease in performance. However,
there are several cases where models exhibiting a similar  parameters count perform very differently, and even cases where models with less parameters perform significantly better. All this indicates that the three parameters $B$, $R$, and $C$  affect performance differently. Another interesting outcome to observe  is that the performance starts decreasing evidently only when employing models with less than 1500K parameters, suggesting that the model size can be easily reduced to 1/5 without major performance drops.

The comments above are quantitatively clear in Table~\ref{tab:results}, which reports the overall results obtained with each configuration, sorted by the total count of the model parameters.  The best results overall are shown in bold font. Model exhibiting similar parameter counts, but very different separation performance are grouped for comparison. For example, the triplet $B=6, R=3, C=64$, to which corresponds $\#params=972$, gives significantly higher SI-SDR than the triplet $B=2, R=2, C=512$, to which corresponds $\#params=1000$. Other interesting cases are the two models with 1400 parameters((8,3,128);(2,3,512)), whose performance differs by more than 3 dB and the three models with 520 parameters ((6,2,64),(4,3,64),(2,3,128)).
\begin{table}[htbp]
\centering
\caption{Separation performance considering different reduced configurations of the Conv-Tasnet baseline (first row). B: number of residual blocks; R: number of repetitions; C: number of channels in depthwise 1D conv layers.}

    \begin{tabular}{|ccc|c|c|}
    \hline
        B & R & C & \# params (K) &SI-SDR (dB)\\
        \hline
8&	3&	512	&5100&	{\bf 12.52}\\
6&	3&	512	&3800&	{\bf 12.80}\\
\hline
6&	2&	512	&2600&	12.02\\
4&	3&	512	&2600&	11.28\\
\hline
8&	1&	512	&1800&	10.36\\
4&	2&	512	&1800&	10.56\\
\hline
8&	3&	128	&1400&	11.92\\
6&	1	&512	&1400	&9.77\\
2&	3&	512	&1400&	8.62\\
\hline
6&	3&	128	&1100&	11.07\\
2&	2&	512	&1000&	7.41\\
6&	3&	64	&972&	10.28\\
\hline
8&	3&	64	&825&	10.26\\
6&	2&	128	&821&	10.27\\
4&	3&	128	&821&	9.93\\
\hline
8&	1&	128	&619&	9.24\\
\hline
6&	2&	64	&520&	9.18\\
4&	3&	64	&520&	8.85\\
2&	3&	128	&518&	7.50\\
\hline
8&	1&	64	&418&	8.31\\
2&	2&	128	&417&	6.49\\
\hline
6&	1	&64  &367&	7.64\\
2&	3&	64	&367&	6.77\\
2&	2&	64	&316&	5.90\\      
\hline
    \end{tabular}
    \label{tab:results}
\end{table}
%
%
 %

Observing the cases above and the general trends in Table~\ref{tab:results}, it appears evident that the  parameter with the major impact on separation performance is the number of residual blocks $B$, in particular with respect to $R$. The motivation is related to the fact that dilation is exponentially increasing in the successive residual blocks. Using a smaller number of blocks limits the dilation and, therefore, the temporal size of the receptive fields. Conversely, reducing the number of channels $C$, in particular for large values of $B$ and $R$, allows reducing considerably the model size, limiting at the same time performance degradation.

\subsection{Extra-dilation}

Using a larger dilation step represents a possible work-around to narrow the effect of curtailment of $B$, since dilation doesn't increase the number of model parameters while, at the same time, it  has a minor effect on the computational requirements. 

Table~\ref{tab:dilation} reports the separation performance of our smallest models when using larger bases for the exponential dilation step, i.e. 4 and 8. When $B=4$   using $\delta=4^{i-1}$ (being $i$ the index of the separation block) 
leads to performance very close to the baseline (first row of the table). In an analogous way, when $B=2$ using $\delta=8^{i-1}$ gives the best performance, although still quite far from the baseline. These results confirm the observation made above on the impact of $B$ on the receptive fields. Note that applying $\delta=8^{i-1}$ to $B=4$ leads to a performance deterioration, which is reasonable because the values  $B=8$ and $\delta=2^{i-1}$ have been specifically optimized on this task. For comparison, the last row reports the performance of SudoRM-RF on the same task. The scaled and extra-dilated version of Conv-Tas-Net with a similar size performs slightly worse than SudoRM-RF but with a higher degree of scalability.

\begin{table}[htbp]
\centering
\caption{Separation performance of the smallest models when different dilations are used. The upper row refers to the baseline, the bottom row reports the SudoRM-RF performance for comparison.}
\begin{tabular}{|ccc|c|c|c|c|}
\hline
\multirow{2}{*}{B}& \multirow{2}{*}{R} &\multirow{2}{*}{C} & \# params &\multicolumn{3}{c|}{SI-SDR (dB)}\\
 & & & (K) &$\delta=2^{i-1}$&$\delta=4^{i-1}$&$\delta=8^{i-1}$\\
\hline
8&3&512&5100& 12.52 & - & - \\
\hline
4&	3&	512&2600&	11.28&	{\bf 12.02}	&	11.10\\
4&	3&	128&821&	 9.90&	{\bf 10.17}	&	 9.66\\
4&	3&	64 &520&	 8.85&	{\bf 9.21}	&	 8.51\\
\hline
2&	3&	512&1400&	 8.62&	8.94	&	{\bf 9.44}\\
2&	3&	128&518&	 7.50&	7.83	&	{\bf 8.02}\\
2&	3&	64 &367&	 6.77&	7.09	&	{\bf 7.26}\\  
\hline
\hline
\multicolumn{3}{|c|}{SudoRM-RF} & $\approx 2660$ & \multicolumn{3}{c|}{13.13}\\
\hline
\end{tabular}
\label{tab:dilation}
\end{table}

\subsection{Knowledge Distillation}
In an attempt to further enhance the performance of the smallest models, we investigated the use of~\ac{KD} from the baseline model, without success. We considered multiple distillation strategies using both the signals and the masks generated by the teacher, exploiting both  SI-SDR and MSE losses. However, we only observed a quicker model convergence during training without a clear benefit in terms of the final performance.

\section{Conclusion and future work}
\label{sec:page}
This paper investigates the impact of scaling approaches on a Conv-Tasnet architecture towards scalable source separation models for on-device deployment. We experimentally assess on the impact of 3 scaling parameters that lead to smaller version of the Conv-Tas-Net model and provide guidelines for practitioners in the field to simplify the compact model design given a target platform. We have shown that the number of residual blocks in the Conv-Tasnet architecture has the major impact on the separation performance and that a possible solution to remediate to the corresponding performance drop is  the adoption of a larger dilation step in the convolution.

Future works will further investigate the use of \ac{KD} or other training strategies as well as architectural modification that could alleviate the detrimental effect of the model compression on the separation performance.

\ninept
\bibliographystyle{IEEEbib}
\bibliography{strings}

\end{document}